\documentclass[%
 aip,
 jcp,%
 amsmath,amssymb,
 reprint,%
citeautoscript,
]{revtex4-1}

\usepackage{graphicx}
\usepackage{dcolumn}
\usepackage{bm}
\usepackage{braket}
\usepackage{mathtools}

\usepackage[usenames,dvipsnames]{color}

\begin{document}

\title{Capturing Vacuum Fluctuations and Photon Correlations in Cavity Quantum Electrodynamics with Multi-Trajectory Ehrenfest Dynamics}

\author{Norah M. Hoffmann}
\affiliation{Max Planck Institute for the Structure and Dynamics of Matter and Center for Free-Electron Laser Science, 22761 Hamburg, Germany}
\author{Christian Sch\"afer}
\affiliation{Max Planck Institute for the Structure and Dynamics of Matter and Center for Free-Electron Laser Science, 22761 Hamburg, Germany}
\author{Angel Rubio}
\affiliation{Max Planck Institute for the Structure and Dynamics of Matter and Center for Free-Electron Laser Science, 22761 Hamburg, Germany}
\affiliation{Center for Computational Quantum Physics (CCQ), Flatiron Institute, 162 Fifth avenue, New York NY 10010}
\author{Aaron Kelly}\email[ ]{aaron.kelly@dal.ca}
\affiliation{Max Planck Institute for the Structure and Dynamics of Matter and Center for Free-Electron Laser Science, 22761 Hamburg, Germany}
\affiliation{Department of Chemistry, Dalhousie University, Halifax, Canada, B3H 4R2}
\author{Heiko Appel}\email[ ]{heiko.appel@mpsd.mpg.de}
\affiliation{Max Planck Institute for the Structure and Dynamics of Matter and Center for Free-Electron Laser Science, 22761 Hamburg, Germany}

\date{\today}

\begin{abstract}

We describe vacuum fluctuations and photon-field correlations in interacting quantum mechanical light-matter systems, by generalizing the application of mixed quantum-classical dynamics techniques. We employ the multi-trajectory implementation of Ehrenfest mean field theory, traditionally developed for electron-nuclear problems, to simulate the spontaneous emission of radiation in a model quantum electrodynamical cavity-bound atomic system. We investigate the performance of this approach in capturing the dynamics of spontaneous emission from the perspective of both the atomic system and the cavity photon field, through a detailed comparison with exact benchmark quantum mechanical observables and correlation functions. By properly accounting for the quantum statistics of the vacuum field, while using mixed quantum-classical (mean field) trajectories to describe the evolution, we identify a surprisingly accurate and promising route towards describing quantum effects in realistic correlated light-matter systems.
  
\end{abstract}

\date{\today}

\maketitle

\section{Introduction}\label{sec:intro}
Profound changes in the physical and chemical properties of material systems can be produced in situations where the quantum nature of light plays an important role in the interaction with the system \cite{Feist2015,Schachenmayer2015,Cirio2016}. A few notable recent examples of such effects are few-photon coherent nonlinear optics with single molecules \cite{maser2016}, direct experimental sampling of electric-field vacuum fluctuations \cite{riek2015,moskalenko2015}, multiple Rabi splittings under ultra-strong vibrational coupling \cite{george2016}, exciton-polariton condensates \cite{byrnes2014,kasprzak2006}, and frustrated polaritons \cite{Schmidt2016}. These exciting developments have been strongly driven by experimental efforts, thus exposing the immediate need for the development and improvement of theoretical approaches that can bridge the gap between quantum optics and quantum chemistry \cite{ruggenthaler2018}.

Due to the similarity of the electron-photon and the electron-nuclear problems, simulation methods that have traditionally been of use in the quantum chemistry community, such as semiclassical and mixed quantum-classical methods, offer a potentially interesting avenue to bridge this gap. In particular, the family of trajectory-based quantum-classical methods has the advantage of providing a very intuitive, qualitative understanding of nonadiabatic molecular dynamics. Further, these techniques typically do not exhibit the pernicious exponential scaling of computational effort inherent in grid-based quantum calculations\cite{Thoss2004}. Available techniques in this family of exact and approximate approaches are Ehrenfest mean field dynamics, fully linearized and partially linearized path integral methods, forward-backward trajectory methods \cite{Hsieh2012,he12,SKR18}, and trajectory surface-hopping algorithms \cite{KM13}. All these techniques have some ability to describe essential quantum mechanical effects such as tunnelling, interference, zero-point energy conservation.

Recently, Subotnik and co-workers have performed investigations of light-matter interactions where an adjusted Ehrenfest theory based method is used to simulate spontaneous emission of classical light \cite{CLSNS18,CLSNS218,LNSMCS18}. Here, in contrast with these works, we focus on the description of quantized light fields. We then generalize the well established multi-trajectory Ehrenfest method to treat quantum mechanical light-matter interactions. We highlight the possibilities and theoretical challenges of this method in comparison to the exact treatment of the quantum system, by applying this approach to investigate spontaneous emission for a model atom in an optical cavity. 

The remainder of this work is divided into three sections: in Sec.\ref{sec:theory} we briefly review general interacting light-matter systems, and the multi-trajectory Ehrenfest dynamics method. In this framework, we then introduce a one-dimensional model system comprising a single (two or three level) atomic system coupled to a multi-mode quantum electrodynamical (QED) cavity. In Sec.\ref{sec:results} we investigate the performance of multi-trajectory Ehrenfest (MTEF) dynamics in describing the process of spontaneous emission. We conclude our results in Sec.\ref{sec:conc} and discuss some prospects for future work.

\section{Theory}\label{sec:theory}
\subsection{Quantum Mechanical Light-Matter Interactions}
To begin, we describe a general coupled field-matter system using Coulomb gauge and the dipole approximation \cite{Faisal1987,Flick2017a}. The total Hamiltonian for the system is \cite{Tokatly2013,Flick2017,Pellegrini2015,Flick2015,Craig1998}
\begin{equation}
\hat{H} = \hat{H}_{A} + \hat{H}_{F} + \hat{H}_{AF}.
\label{G9}
\end{equation}
The first term, $\hat{H}_{A}$, is the atomic Hamiltonian, which may be generally expressed in the spectral representation,
\begin{equation}
\hat{H}_{A} = \sum_k \varepsilon_k | k \rangle \langle k |.
\end{equation}
Here $\{\varepsilon_k,|k\rangle\}$ are the atomic energies and stationary states of the atomic system in absence of coupling to the cavity. The second term is the Hamiltonian of the uncoupled cavity field $\hat{H}_{F}$,
\begin{equation}
\hat{H}_{F} = \frac{1}{2}\sum_{\alpha = 1}^{2N} \left(\hat{P}^{2}_{\alpha} + \omega^{2}_{\alpha}\hat{Q}_{\alpha}^{2} \right).
\label{G12}
\end{equation} The photon-field operators, $\hat{Q}_{\alpha}$ and $\hat{P}_{\alpha}$, obey the canonical commutation relation, $[\hat{Q}_{\alpha},\hat{P}_{\alpha'}] = \imath\hbar\delta_{\alpha,\alpha'}$, and can be expressed using creation and annihilation operators for each mode of the cavity field,
\begin{eqnarray}
\hat{Q}_{\alpha} &=& \sqrt{\frac{\hbar}{2\omega_{\alpha}}}(\hat{a}^{+}_{\alpha} + \hat{a}_{\alpha}), \\
\hat{P}_{\alpha} &=& i\sqrt{\frac{\hbar\omega_{\alpha}}{2}}(\hat{a}^{+}_{\alpha} - \hat{a}_{\alpha}),
\end{eqnarray} where $\hat{a}^\dagger_\alpha$ and $\hat{a}_\alpha$ denote the usual photon creation and annihilation operators for photon mode $\alpha$. The coordinate-like operators, $\hat{Q}_{\alpha}$, are directly proportional to the electric displacement operator, while the conjugate momenta-like operators, $\hat{P}_{\alpha}$, are related to the magnetic field \cite{Pellegrini2015,Flick2015}. The upper limit of the sum in Eq.(\ref{G12}) is $2N$, as there are (in principle) two independent polarization degrees of freedom for each photon mode, however in the 1D cavity models presented here only a single polarization will be considered. 

The final term in Eq.(\ref{G9}) represents the coupling between the atom and the cavity field,
\begin{equation} 
\hat{H}_{AF} = \sum_{\alpha=1}^{2N}\Big(\omega_{\alpha}\hat{Q}_{\alpha}(\lambda_{\alpha}\cdot \hat{\mu}) + \frac{1}{2}\Big(\lambda_{\alpha} \cdot \hat{\mu} \Big)^2 \Big),
\end{equation} where we denote ${\hat{\mu}}$ as the electronic dipole moment vector of the atomic system, and ${\lambda}_{\alpha}$ as the electron-photon coupling vector \cite{Tokatly2013,Flick2015}. In the case of a two-level electronic system the quadratic term in the atom-field coupling Hamiltonian simply results in a constant energy shift and hence has no effect on observables \cite{SRR18}, and we neglect this term in the case of the three level model system. Furthermore, we note that this Hamiltonian can easily be extended to include nuclear degrees of freedom, however this has been omitted in the present work.

\begin{figure}
\includegraphics[width=0.5\linewidth]{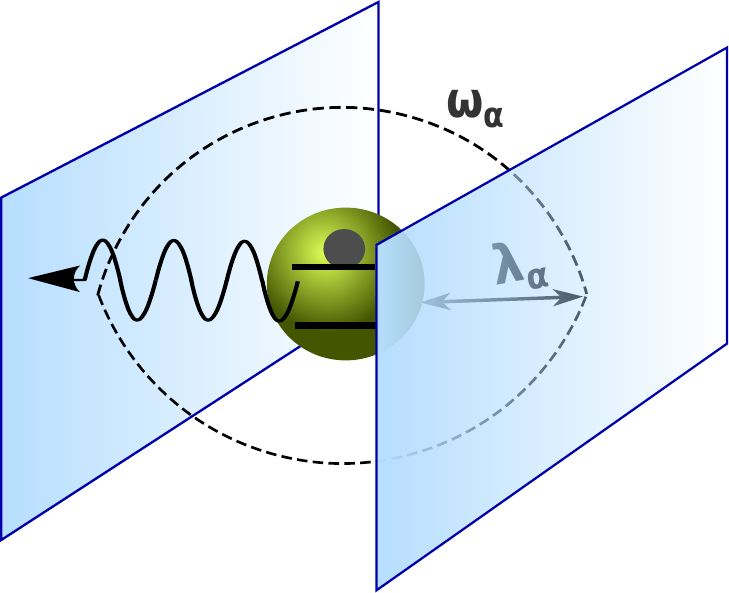}
\caption{Model atomic system in an electromagnetic cavity: Atom (green) trapped between two mirror-like surfaces of the cavity, supporting $2N$ photon modes with frequencies $\omega_{\alpha} = \frac{2\pi \hbar c \alpha}{L}$, where $\alpha = \{1,2, ...,2N\}$ and $L$ is the distance between the mirrors. The strength of the interactions between each mode of the cavity field and the atomic system is $\lambda_{\alpha}$.} \label{AB1}
\end{figure}

\subsection{Multi-Trajectory Ehrenfest Dynamics}
In this section we apply the well-known multi-trajectory Ehrenfest method, traditionally introduced to study electron-nuclear systems \cite{Ehrenfest1927,Makri1987,C.Tully1998}, to coupled light-matter systems\cite{McLachlan1964,C.Tully1998,Flick2017a}. 

A particularly simple and instructive route to derive the MTEF mean field theory is via the quantum-classical Liouville (QCL) equation\cite{qcle}. This equation of motion for the density matrix is formally exact for an arbitrary quantum mechanical system that is bilinearly coupled to a harmonic environment, as is the case in the atom-field Hamiltonian studied here. The QCL equation can be written in a compact form as
\begin{eqnarray} \label{eq:qcle}
  && \frac{\partial }{\partial t}\hat{\rho}_W (X, t) =-i {\mathcal L}\hat{\rho}_W (X,t).
\end{eqnarray}
It describes the time evolution of $\hat{\rho}_W(X,t)$, which is the partial Wigner transform of the density operator taken over the photon field coordinates, which are thus represented by continuous phase space variables $X=(Q,P)=(Q_1,Q_2,...,Q_{2N},P_1,P_2,...,P_{2N})$.  The partial Wigner transform of the density operator, $\hat{\rho}$, is defined as
\begin{eqnarray} \label{eq:wigner}
\hat{\rho}_{W}(Q,P) = \frac{1}{(2\pi\hbar)^{2N}}\int dZ e^{i P \cdot Z} \langle Q - \frac{Z}{2} | \hat{\rho} | Q +\frac{Z}{2}\rangle.  
\end{eqnarray}
The QCL operator is defined as
\begin{equation}\label{eq:qcl_op}
i{\mathcal L} \cdot = \frac{i}{\hbar}[\hat{H}_W,\cdot] - \frac{1}{2}(\{\hat{H}_W,\cdot\}
-\{\cdot,\hat{H}_W\}),
\end{equation}
where $\hat{H}_{W}$ denotes the Wigner transform of $\hat{H}$, $[\cdot,\cdot]$ is the commutator, and $\{\cdot,\cdot\}$ is the Poisson bracket in the phase space of the environmental variables. 

At this point, one may arrive at MTEF equations by assuming that the total density of the system can be written as an uncorrelated product of the atomic and photonic reduced densities at all times,
\begin{equation}
\hat{\rho}_W(X,t)=\hat{\rho}_A(t) \rho_{F,W}(X,t),
\end{equation} 
where the reduced density matrix of the atomic system is
\begin{equation}\label{eq:rdm}
\hat{\rho}_{A} (t) = \text{Tr}_F \Big( \hat{\rho}(t) \Big) = \int dX \hat{\rho}_W (X,t),
\end{equation} and the Wigner function of the cavity field is $\rho_{F,W}(X, t) = Tr_A ( \hat{\rho}_W (X,t))$. 
If one seeks solutions to the QCL equation of this form, the Ehrenfest mean-field equations of motion for the atomic system are obtained:
\begin{equation}
    \frac{d}{dt} \hat{\rho}_A(t) = -i\Big[ \hat{H}_A + \hat{H}_{AF,W}(X(t)), \hat{\rho}_A(t)\Big],
\end{equation}
where $\hat{H}_{AF,W}$ denotes the Wigner transform of $\hat{H}_{AF}$. The evolution of the Wigner function of the photon field can be represented as a statistical ensemble of independent trajectories, with weights $w^j$, \begin{equation}
\rho_{F,W} (X,t) = \sum_j^{N_{traj}} w^j \delta(X-X^j(t))
\end{equation} that evolve according to Hamilton's equations of motion,
\begin{equation}
    \frac{\partial Q_{\alpha}}{\partial t} =  \frac{\partial H_{F,W}^{Eff}}{\partial P_{\alpha}},\quad \frac{\partial P_{\alpha}}{\partial t} = - \frac{\partial H_{F,W}^{Eff}}{\partial Q_{\alpha}}.
\end{equation}
The effective photon field Hamiltonian is, 
\begin{equation}
    H^{Eff}_{F,W} = \frac{1}{2}\sum_{\alpha}\Big( P_{\alpha}^2 + \omega_{\alpha}^2 Q_{\alpha}^2 - 2 \omega_{\alpha} \lambda_{\alpha} Q_{\alpha}  \mu (t)   \Big),
\end{equation}
where $ \mu (t)  = \text{Tr}_A(\hat{\rho}_A(0) \hat{\mu}(t))$.

The exact expression for the average value of any observable, $\langle O (t) \rangle $, can be written as
\begin{eqnarray} 
\langle O (t) \rangle  = \text{Tr}_A \int dX  \hat{O}_W(X,t) \hat{\rho}_W(X,t=0).
\end{eqnarray} 

We note here that for this class of systems the Ehrenfest equations of motion for the photon field coordinates correspond to a mode resolved form of Maxwell's equations. In applying the MTEF dynamics method numerically, we use the above expressions in the following manner: 
\begin{enumerate}
    \item We first perform Monte Carlo sampling from the Wigner transform of the initial density operator of the photon field $\hat{\rho}_{F,W}(X,0)$ to generate an ensemble of initial conditions, for the trajectory ensemble $(Q_{\alpha}^j(0),P_{\alpha}^j(0))$. In this work we used uniform weights $w^j = \frac{1}{N_{traj}}$, however other importance sampling schemes could be employed as the only requirement is that the sum of the weights is normalized, $\sum_j w^j = 1$. 
    \item We evolve each initial condition independently according to the Ehrenfest equations of motion, producing a trajectory. In the following we refer to such a solution as an ensemble of independent trajectories. 
    \item Average values are constructed by summing over the entire trajectory ensemble, and normalizing the result with respect to $N_{traj}$, the total number of trajectories in the ensemble, \\
$\langle O (t) \rangle  =  \sum_j^{N_{traj}} \text{Tr}_A \Big( \hat{O}_W(Q^j,P^j,t)\hat{\rho}_A(0)\Big)/N_{traj}$.
\end{enumerate}

Here $\rho_{F,W}(X,0)$ is Wigner transform of the zero temperature vacuum state \begin{equation}
\rho_{F,W}(X,0)=\prod_{\alpha} \frac{1}{\pi}\exp{\left[-\frac{P^{2}_{\alpha}}{\omega_{\alpha}} - \omega_{\alpha} Q_{\alpha}^2\right]}.
\label{G20}
\end{equation}

\subsection{Observables and Normal Ordering}

Before we proceed with a discussion of our simulation results, we must note that the Wick normal ordered form for operators (denoted $:\hat{O}:$ for some operator $\hat{O}$) is used when calculating average values in this study. The reason for using the normal ordered form, in practice, is to remove the effect of vacuum fluctuations from the results, which ensures that both $\langle E \rangle = 0$ and $\langle I \rangle = 0$, irrespective of the number of photon modes in the cavity field, when the field is in the vacuum state. The effect of this operator ordering is particularly evident for the photon number operator,
\begin{equation}
:\hat{N}_{pt}: = \frac{1}{2}\sum_{\alpha}\Big(\frac{\hat{P}^{2}_{\alpha}}{\omega_{\alpha}} + \omega_{\alpha}\hat{Q}^{2}_{\alpha} -1\Big),
\label{G5}
\end{equation} where normal ordering produces a constant shift due to the zero-point energy term.

\begin{figure}[h!]{
\includegraphics[width=\columnwidth]{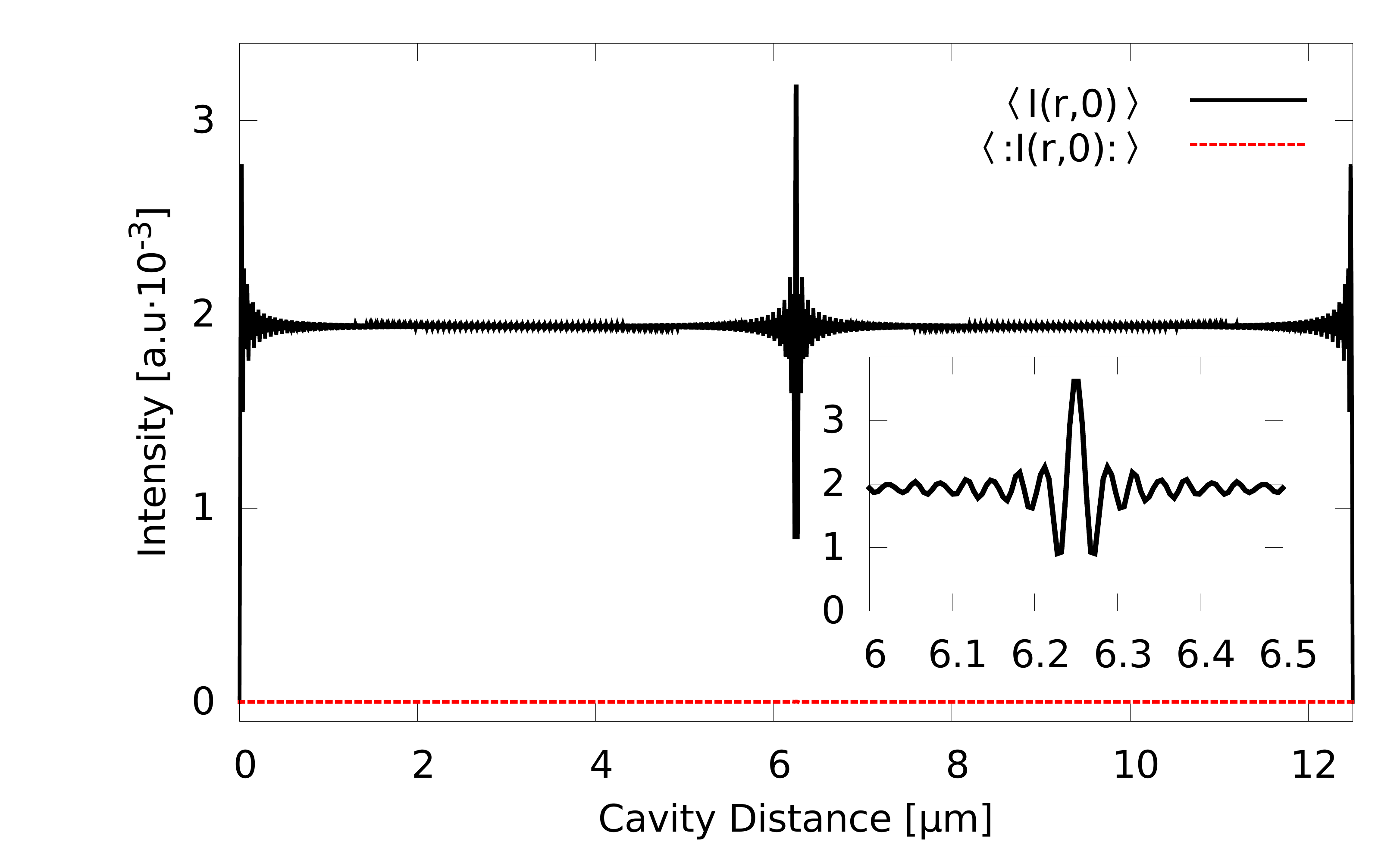}}
\caption{Average value of the cavity electric field intensity. Wick normal ordering has been applied to the operator in the case of the red dashed line, whereas the solid black line corresponds to the original operator. The cavity field is prepared in the vacuum state, at zero temperature.}
\label{AB2}
\end{figure}

The quantized electric field operator is defined as \begin{equation}
\hat{E}(r,t) = \sum_{\alpha}\sqrt{2\omega_{\alpha}} \zeta_{\alpha}(r) \hat{Q}_{\alpha}(t)
\end{equation} with 
\begin{equation}
\zeta_{\alpha}(r) = \sqrt{\frac{\hbar \omega_{\alpha}}{\epsilon_{0} L}} \sin\Big(\frac{\alpha \pi}{L} r\Big).
\end{equation}
The corresponding normal-ordered electric field intensity operator is given by
\begin{equation}
:\hat{E}^2(r,t): = :\hat{I}(r,t): = 2\sum_{\alpha} \omega_{\alpha}\zeta^{2}_{\alpha}(r)\hat{Q}_{\alpha}^{2}(t) - \sum_{\alpha}\zeta^{2}_{\alpha}(r). 
\label{G15}
\end{equation}
The effect of normal ordering on this quantity is shown in Fig.~\ref{AB2}, where the intensity of the electric field is plotted in both its canonical and normal ordered forms. In addition to a constant shift with respect to the normal ordered quantity, which is identically zero, the canonical average field intensity also displays additional oscillations near the boundaries and the atomic position, corresponding to the vacuum fluctuations for this system.

We also consider the second order correlation function for the photon field \cite{Glauber},
\begin{equation}
:g^{2}(r_{1},r_{2},t): = \frac{ \langle : \hat{E}^{+}(r_{1},t) \hat{E}^{+}(r_{2},t)\hat{E}(r_{2},t) \hat{E}(r_{1},t) : \rangle }{ \langle : \hat{I}(r_{1},t) : \rangle \quad \langle : \hat{I}(r_{2},t) : \rangle }.
\label{G14}
\end{equation} This function is frequently used in quantum optics to discriminate between classical light and non-classical states of the photon field that exhibit photon bunching $(g^2 > 1)$ or photon anti-bunching $(g^2 <  1)$. The normal-ordered form of the numerator in $g^2$, also referred to  $G^{2}(r_{1},r_{2},t)$, is
\begin{eqnarray}
&:&G^{2}(r_{1},r_{2}, t): =  4\sum_{\alpha} \omega_{\alpha}^2\zeta_{\alpha}(r_{1})\zeta_{\alpha}(r_{2})\zeta_{\alpha}(r_{2})\zeta_{\alpha}(r_{1})\hat{Q}_{\alpha}^{4}(t)   \nonumber \\ \nonumber 
&-& \sum_{\alpha \beta}\Big(4\zeta_{\beta}(r_{1})\zeta_{\beta}(r_{2})\zeta_{\alpha}(r_{1})\zeta_{\alpha}(r_{2}) +\zeta^{2}_{\beta}(r_{2})\zeta^{2}_{\alpha}(r_{1}) \nonumber \\
& + &\zeta^{2}_{\beta}(r_{1})\zeta^{2}_{\alpha}(r_{2})\Big)\cdot2\omega_{\alpha}\hat{Q}^{2}_{\alpha}(t)  . \label{G19}
\end{eqnarray}

The partial Wigner transforms of the polynomial functions of the bath coordinate operators are simply polynomial functions of the continuous bath coordinates, $\Big(\hat{Q}_{\alpha}^n(t)\Big)_W = (Q_{\alpha}(t))^n$ \cite{Hillery}. The same is also true for the corresponding momenta, and thus the average values of the preceding operators can be easily calculated using mean-field trajectories.

\subsection{Model System}
Following previous work\cite{Buzek,Flick2017}, we investigate a model atomic system in a one-dimensional electromagnetic cavity, as depicted in Fig 1. 
\begin{eqnarray}
\hat{H} = &\sum_{k=1}^m\epsilon_{k}\ket{k}\bra{k} + \frac{1}{2}\sum_{\alpha}^{2N}\Big(\hat{P}^{2}_{\alpha}+ \omega^{2}_{\alpha}\hat{Q}^{2}_{\alpha}\Big) \\ \nonumber +&\sum_{\alpha}^{2N}\sum_{k,l =1}^{m}\mu_{kl}\omega_{\alpha}\lambda_{\alpha}(r_A)\hat{Q}_{\alpha} \ket{k}\bra{l},
\label{G1}
\end{eqnarray}
where the upper limit of the first and last summation $m$ denotes the number of atomic energy levels. In the case of a two-level atomic system, this corresponds to a special case of the spin-boson model. With the position of the atom fixed at $r_A = \frac{L}{2}$ in this study, half of the $2N$ cavity modes decouple from the atomic system by symmetry. We adopt the same parameters as in Ref. \cite{Flick2017,Su1991}, which are based on a 1D Hydrogen atom with a soft Coulomb potential (in atomic units): $\{\varepsilon_1,\varepsilon_2\} = \{-0.6738, -0.2798\}$, $\lambda_{\alpha}(\frac{L}{2}) = 0.0103\cdot(-1)^{\alpha}$, $L = 2.362\cdot 10^{5}$ and $\mu_{12} = 1.034$. For the three-level atom, we adopt all the same parameters for the field and the atom-field coupling as for the two-level case. The atomic energies for the three level model are $\{\varepsilon_1,\varepsilon_2,\varepsilon_3\} = \{-0.6738, -0.2798, -0.1547\}$ and as before the numerical parameters are based on the 1D soft-Coulomb Hydrogen atom. The dipole moment operator only couples adjacent states, such that, the only nonzero matrix elements are $\{\mu_{12},\mu_{23}\} = \{1.034,-2.536\}$ and their conjugates.

\section{Results and Discussion}\label{sec:results}

We now investigate the performance of the MTEF method in the context of cavity-bound spontaneous emission. In all calculations shown below we use 400 photon modes to represent the cavity field. We choose the atom to be initially in the excited state, and the cavity field is in the vacuum state at zero temperature. In all simulations reported here we use an ensemble of $N_{traj} = 10^4$ independent trajectories, sampled from the Wigner transform of the initial field density operator given in the previous section. This level of sampling is sufficient to converge the atomic observables to graphical accuracy, however observables and correlations functions of the photon-field would require a slightly larger trajectory ensemble for graphical convergence. All observables shown below correspond to their normal ordered forms. For our benchmark numerical treatment we solved the time-dependent Schr\"odinger equation by using a truncated Configuration Interaction (CI) expansion. More precisely, the photon field state-space is truncated at two excitations per photon mode, whereas for the atomic system a two and three state discrete variable representation is used in each case. Numerical convergence is checked to ensure that the CI basis that we employ is complete for the models and parameter regimes studied in this work.

\subsection{Two Level Atom: One Photon Emission Process}\label{Se3a}

\begin{figure}[h!]{
\includegraphics[width=\linewidth]{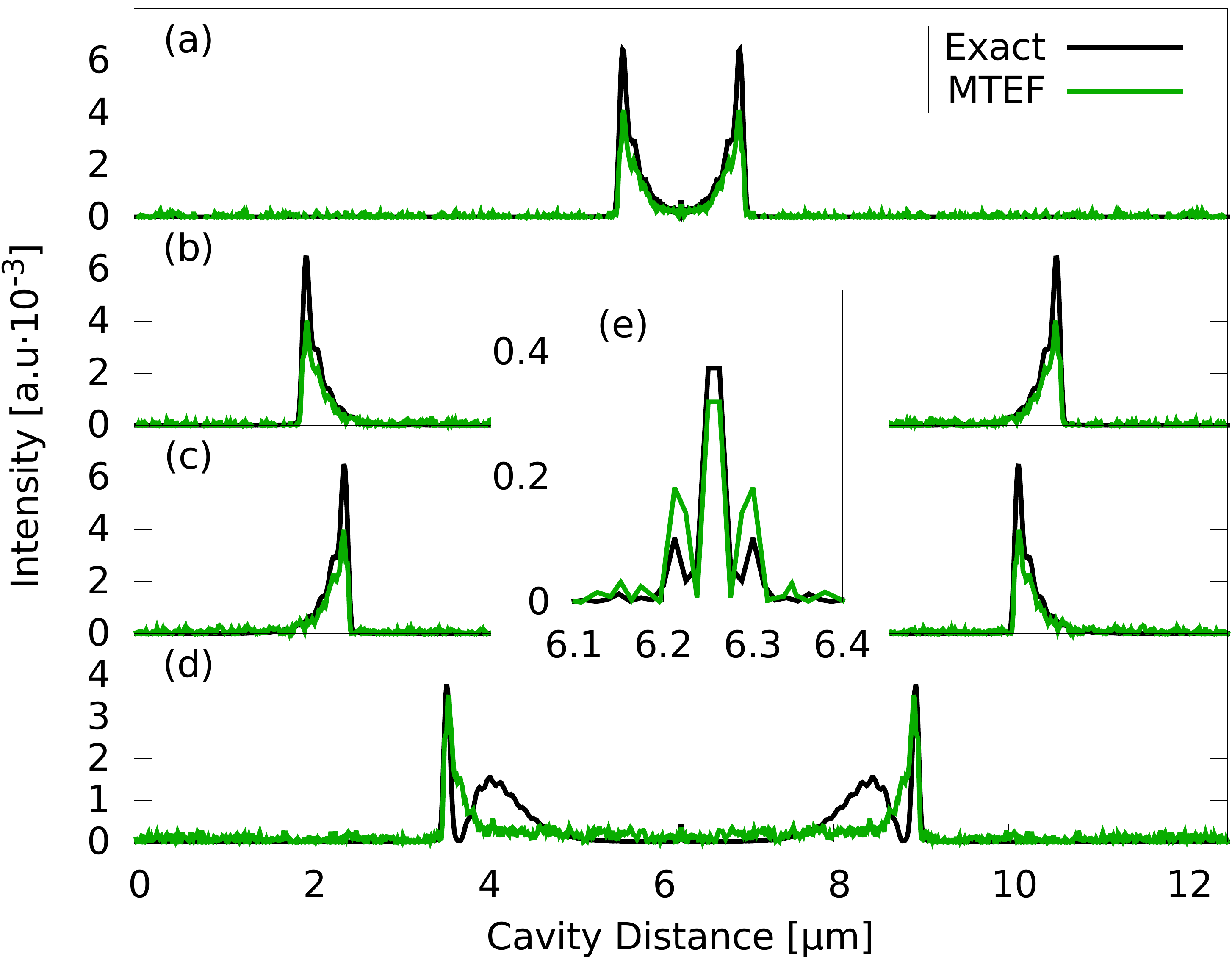}}
\caption{Time-evolution of the average field intensity for the one-photon emission process, at  four different time snapshots: (a) $t = 100 ~a.u.$, (b) $t = 600 ~a.u.$, (c) $t = 1200 ~a.u.$, (d) $t = 2100 ~a.u.$.~(e) Zoom-in of the polariton-peak at the atomic position. Exact simulation results (black) and MTEF dynamics (green).}
\label{AB3}
\end{figure}

\begin{figure}[h!]{
\includegraphics[width=\linewidth]{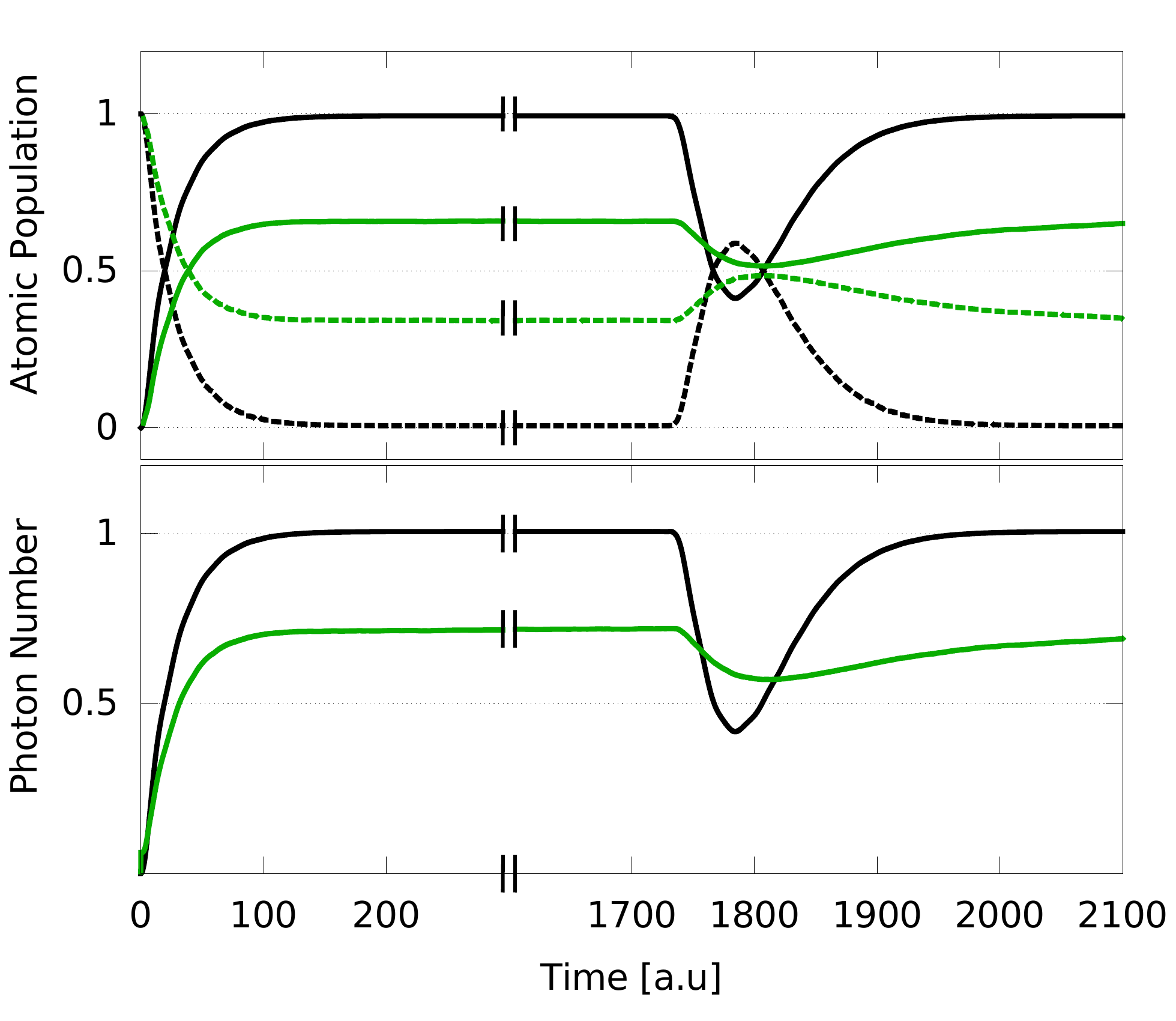}}
\caption{ Time evolution of the atomic state populations (top panel), and the total photon number (bottom panel). Top panel: Solid lines represent the atomic ground state, and dashed lines represent the excited state. Both panels: Exact simulation results (black) and MTEF (green).}
\label{AB4}
\end{figure}

\begin{figure}[h!]{
\includegraphics[width=1\linewidth]{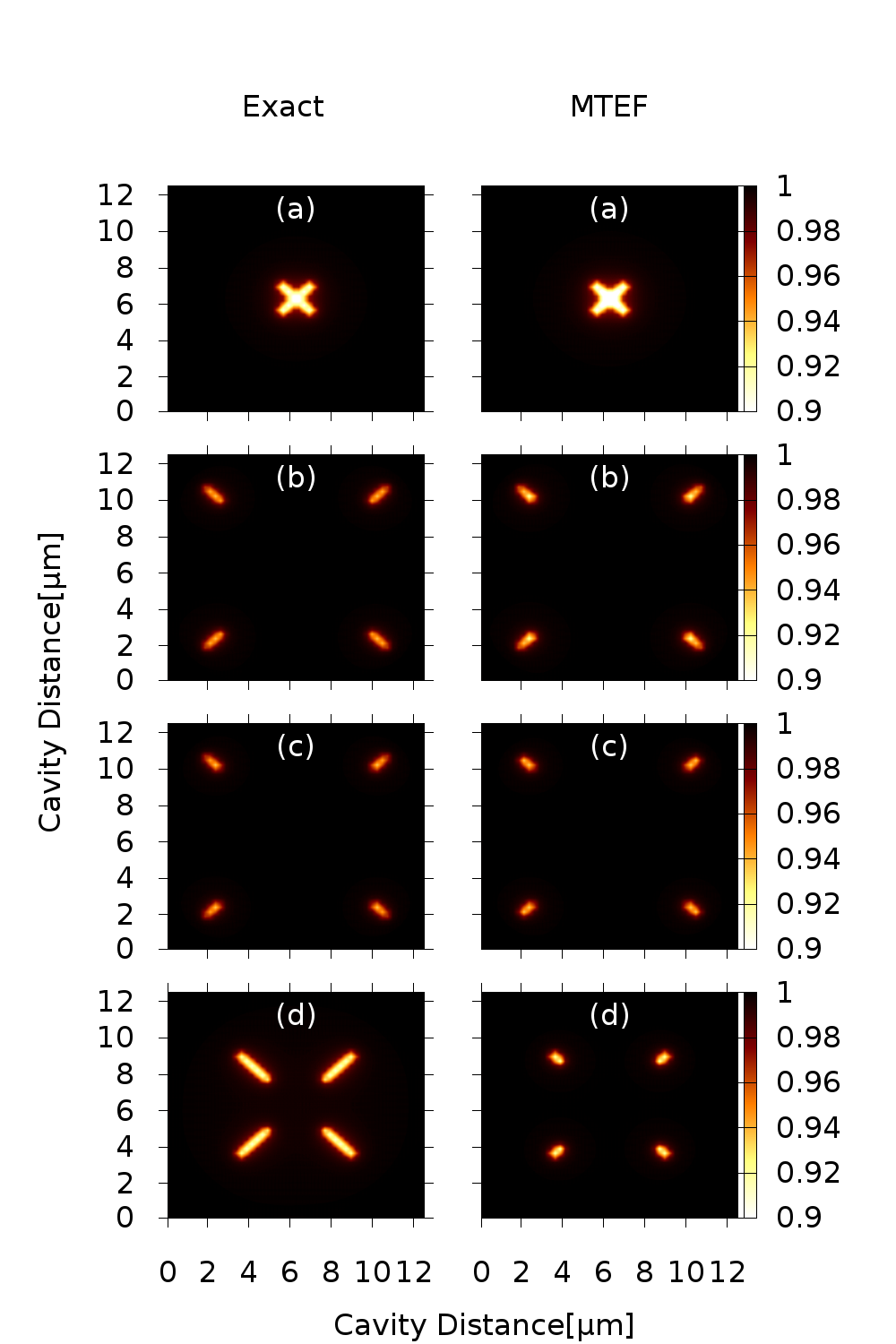}}
\caption{Second order correlation function for the photon field, $g^{2}(r_{1},r_{2},t)$ for the two-level model, plotted at four time snapshots: (a) $t = 100 ~a.u$., (b) $t = 600 ~a.u.$, (c) $t = 1200 ~a.u.$, (d) $t = 2100 ~a.u.$. Exact (left panels), and MTEF (right panels).}
\label{AB5}
\end{figure}

\begin{figure}[h!]{
\includegraphics[width=\linewidth]{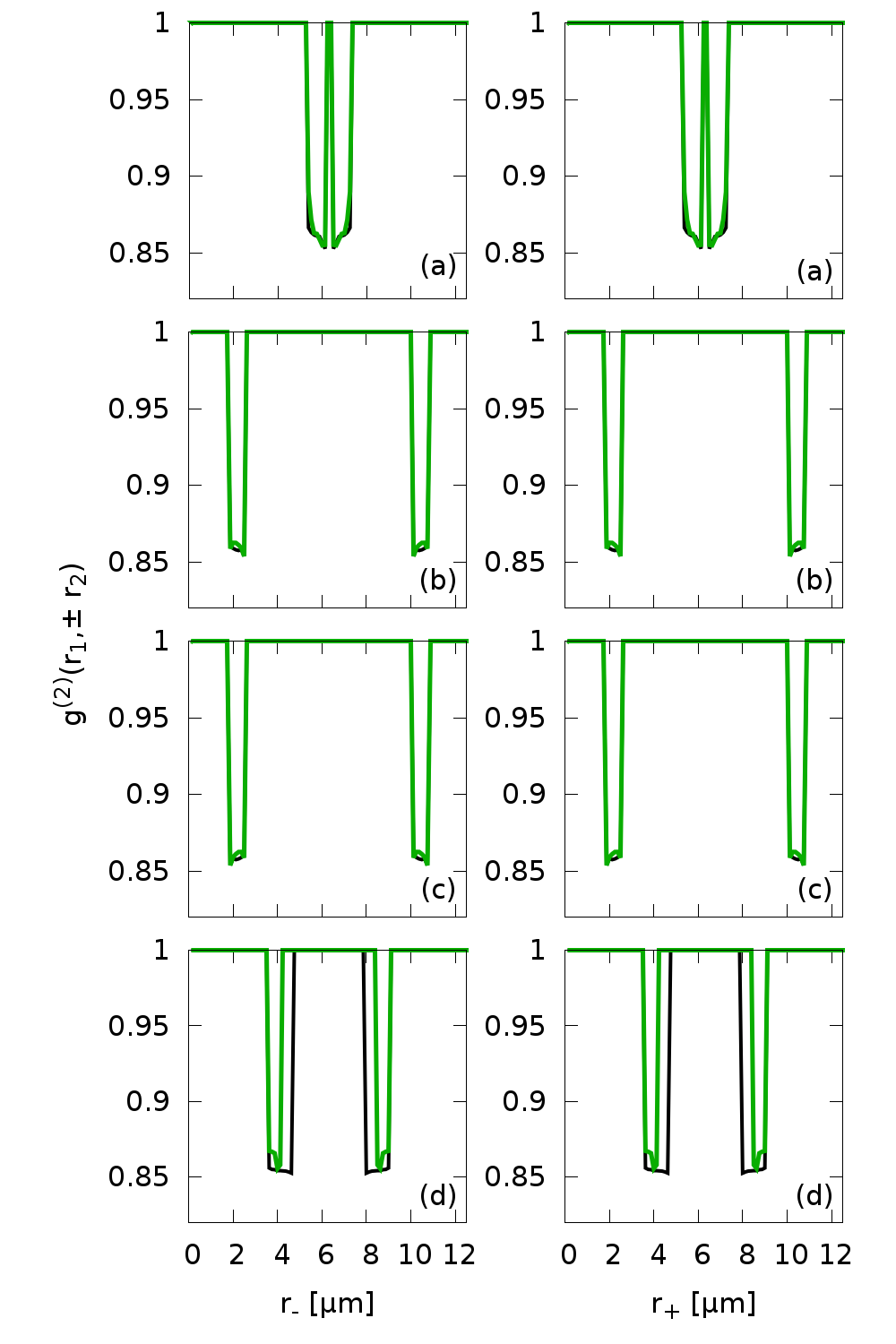}}
\caption{Associated one-dimensional diagonal cuts $g^2(r_{\pm},t)$ of the second order correlation function, exact (black) and MTEF (green), plotted at four time snapshots:(a) $t = 100 ~a.u.$, (b) $t = 600 ~a.u.$, (c) $t = 1200 ~a.u.$, (d) $t = 2100 ~a.u.$.}
\label{AB6}
\end{figure}

In Fig.~\ref{AB3} we show the intensity of the cavity field along the axis of the cavity, at four different times. As the spontaneous emission process proceeds, a photon wavepacket with a sharp front is emitted from the atom and travels toward the boundaries where it is reflected, and then travels back to the atom (e.g. panel (c) of Fig.~3). The emitted photon is then absorbed and re-emitted by the atom, which results in the emergence of interference phenomena in the electric field. This produces a photonic wavepacket with a more complex shape (e.g. panel (d) of Fig.~3). We observe that the MTEF simulations capture the qualitative character of the spontaneous emission process extremely accurately, as well as the wavepacket propagation through the cavity. However, MTEF dynamics fails to reproduce the interference phenomena in the field due to re-emission. We do note, however, that the MTEF simulations are capable of describing the remaining field intensity at the atomic position (e.g panel (e) of Fig.~3). This feature corresponds to a bound electron-photon state, or polariton, which is an emergent hybrid state of the correlated light-matter system. 

We also plot the excited state population of the atomic system, and the average value of the photon number for the field, in Fig.~\ref{AB4}. Again, MTEF is able to capture the qualitative behaviour of both of these quantities very nicely. However, it fails to quantitatively reproduce the correct values for the emitted photon number and atomic population transfer, as these quantities are underestimated. Furthermore, as a result of this loss in accuracy, only a part of the subsequent re-excitation and re-emission processes are captured.

In Fig.~\ref{AB5} we investigate the normalized second order correlation function, $g^2(r_1,r_2,t)$ for the cavity photon field. The unperturbed vacuum state, which is coherent, corresponds to $g^{2}(r_{1},r_{2},t) = 1$, given by the black background seen in Fig.~\ref{AB5}. The vacuum state is disturbed by the emitted wavepacket, corresponding to anti-bunched light with $g^{2}(r_{1},r_{2},t)<1$. The simplicity of the one-dimensional/one-photon process is quite clear in Fig.~\ref{AB6}, where we show the associated one-dimensional cuts of $g^2$, along with projections of $g^2(r_1,r_2,t)$ along the positive and negative diagonals, $r_\pm = (r_1\pm r_2)/\sqrt{2}$. Here we find similar to the intensity a nice qualitative agreement between MTEF and the exact result for the first three time snapshots. However for the last time-snap-shot the exact solution shows a broader correlation than MTEF, which corresponds to the fact that MTEF is not able to accurately capture re-emission. Furthermore, as we only consider a one-photon process in this case, the correlation is symmetric in $r_{+}$ and $r_{-}$. 

\subsection{Three Level Atom: Two Photon Emission Process}\label{Se3b}

We now investigate the three-level system, for the same observables as the previous section. The initial state for the atomic system is now the second excited state. The photonic initial state remains the zero temperature vacuum state.

\begin{figure}[h!]{
\includegraphics[width=1\linewidth]{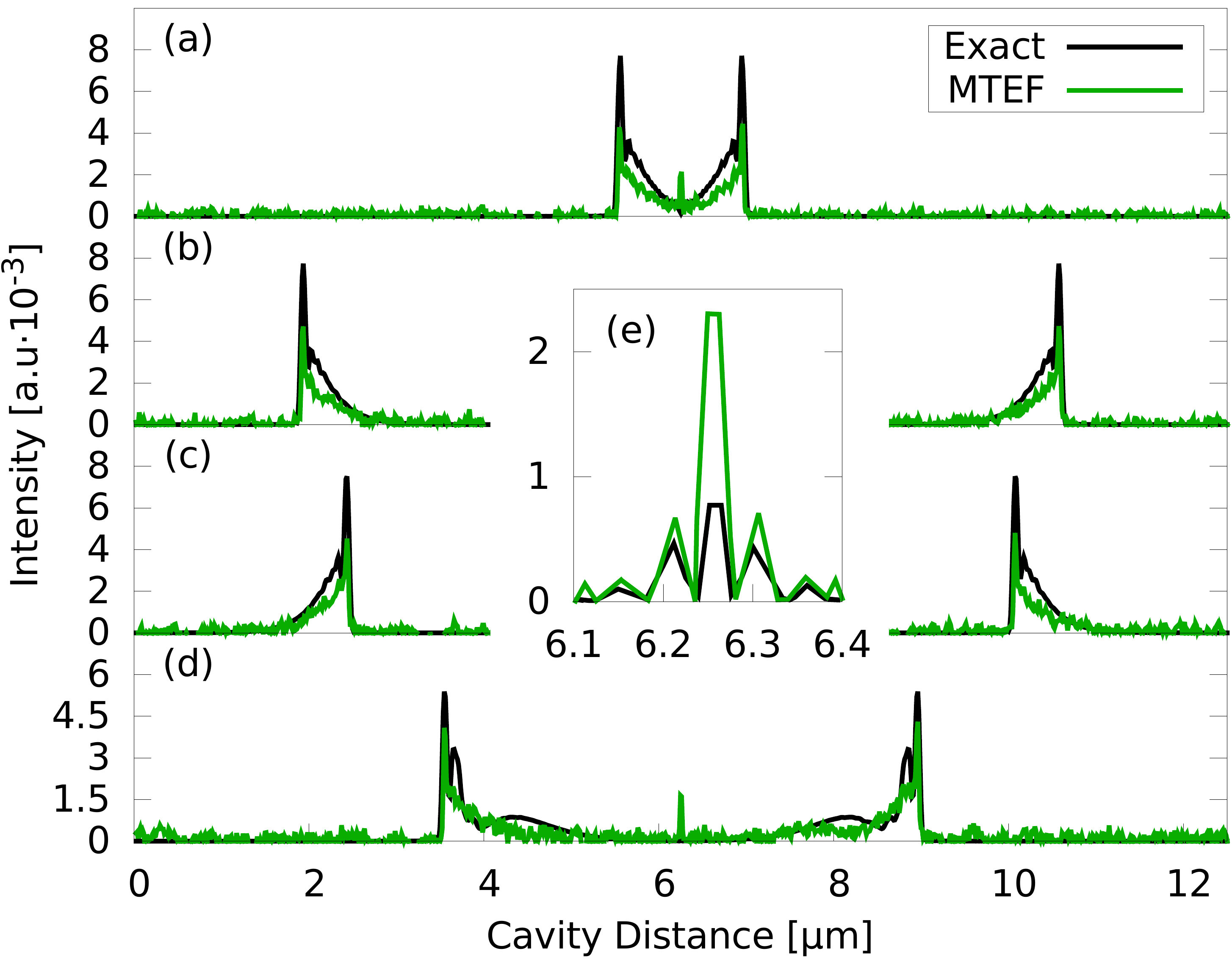}}
\caption{Time-evolution of the average field intensity for the two-photon emission process, at time four different snapshots: (a) $t = 100 ~a.u.$, (b) $t = 600 ~a.u.$, (c) $t = 1200 ~a.u.$, (d) $t = 2100 ~a.u.$.~(e) Zoom-in of the polariton-peak at the atomic position. Exact (black) and MTEF (green).}
\label{AB7}
\end{figure}

In Fig.~\ref{AB7} we show the intensity of the cavity field during the two-photon emission process. Similar dynamics are observed compared with the two-level case. However, due to the additional intermediate atomic state, we now observe a double-peak feature in the emitted photonic wavepacket. This feature corresponds to the emission of two photons, as the excited atom initially drops to the first excited state emitting one photon, and then further relaxes to the ground state, emitting a second photon. The polariton peak (the central feature in the field intensity profile) is overestimated in the MTEF simulations. This overestimation is due to the incomplete relaxation of the second excited state within the Ehrenfest description.

\begin{figure}[h!]{
\includegraphics[width=1\linewidth]{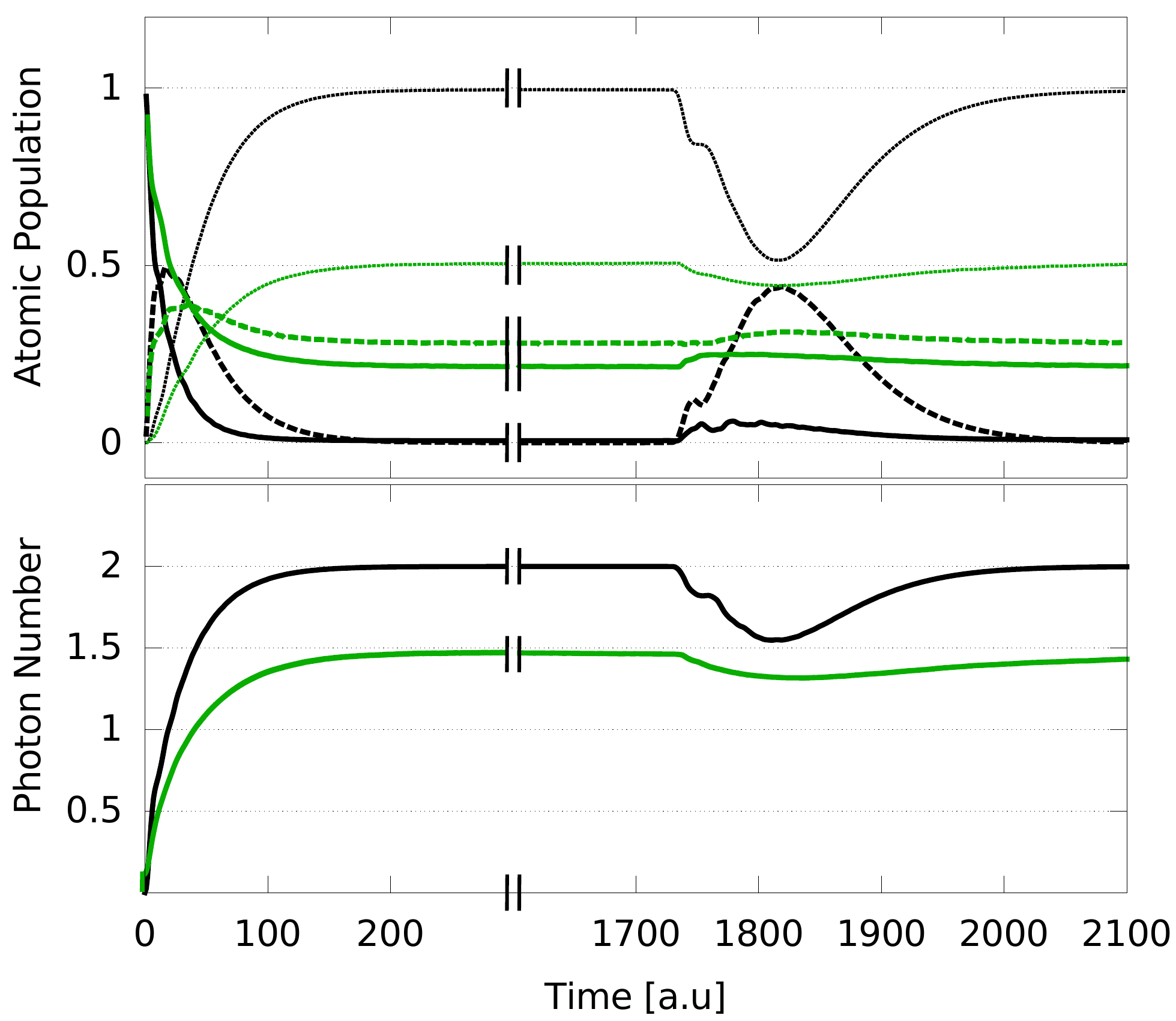}}
\caption{Top panel: Time evolution of the atomic state populations: Solid line ($m=3$), dashed lines ($m=2$), and dotted line ($m=1$). Bottom panel: Total photon number as a function of time. Exact simulation results (black) and MTEF (green).}
\label{AB8}
\end{figure}

In Fig.~\ref{AB8}, we show the time evolution of the atomic state populations and total photon number. Again, the emitted photonic wavepacket moves through the cavity, is reflected at the mirrors, and returns to the atom. The first and second excited state are then repopulated due to stimulated absorption. A second spontaneous emission process ensues, and the emitted field again takes on a more complex profile due to interference. For the intensity, as well as the atomic population and photon number, we observe that MTEF displays qualitatively correct short-time dynamics. However it fails to describe the correct spatial structure of the (re)emitted two photon wavepacket, as well as the correct amplitude for the observables, in accordance with what was observed previously in the two-level case. 

\begin{figure}[h!]{
\includegraphics[width=0.98\linewidth]{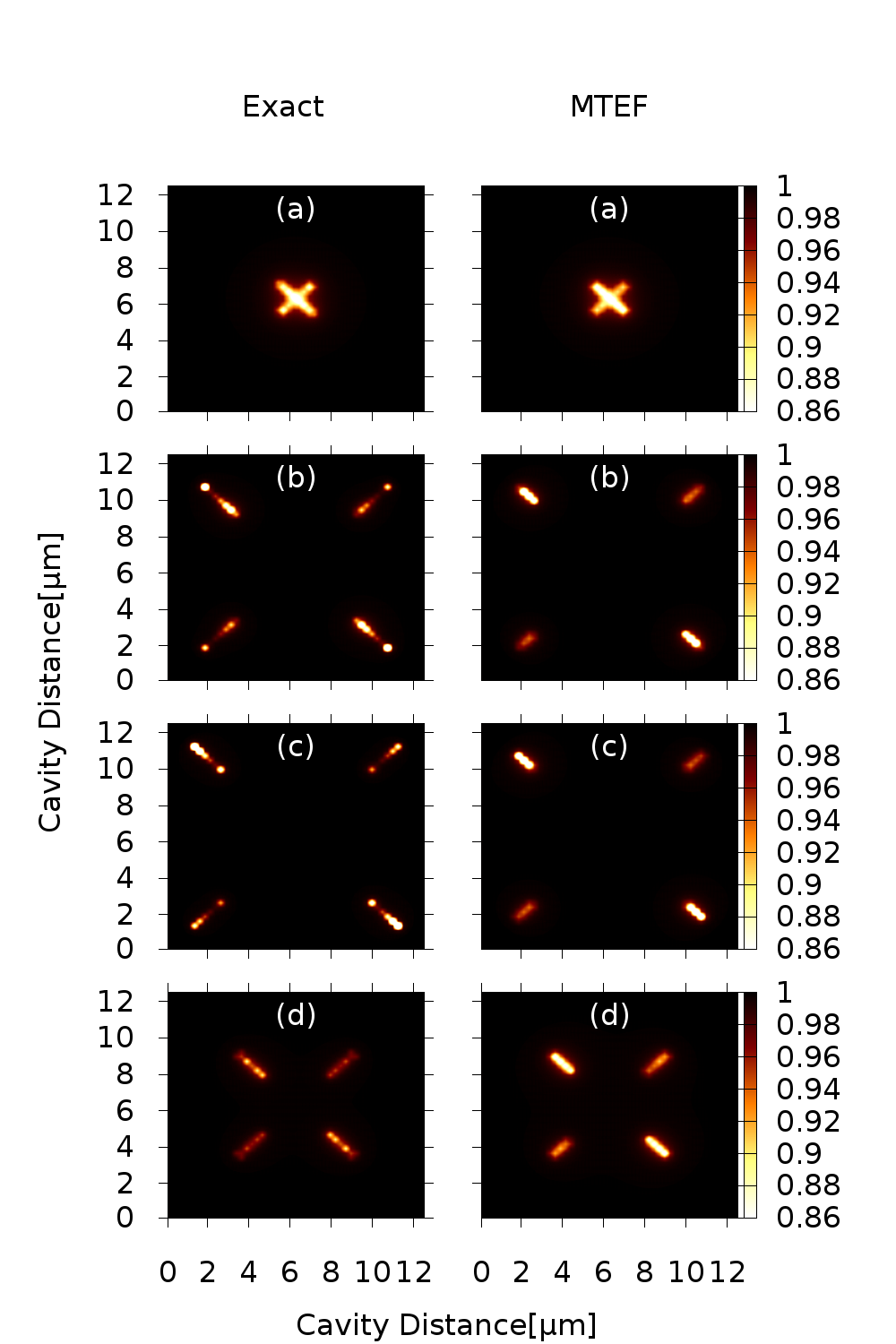}}
\caption{Second order correlation function for the photon field, $g^{2}(r_{1},r_{2},t)$ for the three-level model, plotted at four time snapshots: (a) $t = 100 ~a.u.$, (b) $t = 600 ~a.u.$, (c) $t = 1200 ~a.u.$, (d) $t = 2100 ~a.u.$. Exact (left panels) and MTEF (right panels).}
\label{AB9}
\end{figure}

\begin{figure}[h!]{
\includegraphics[width=1\linewidth]{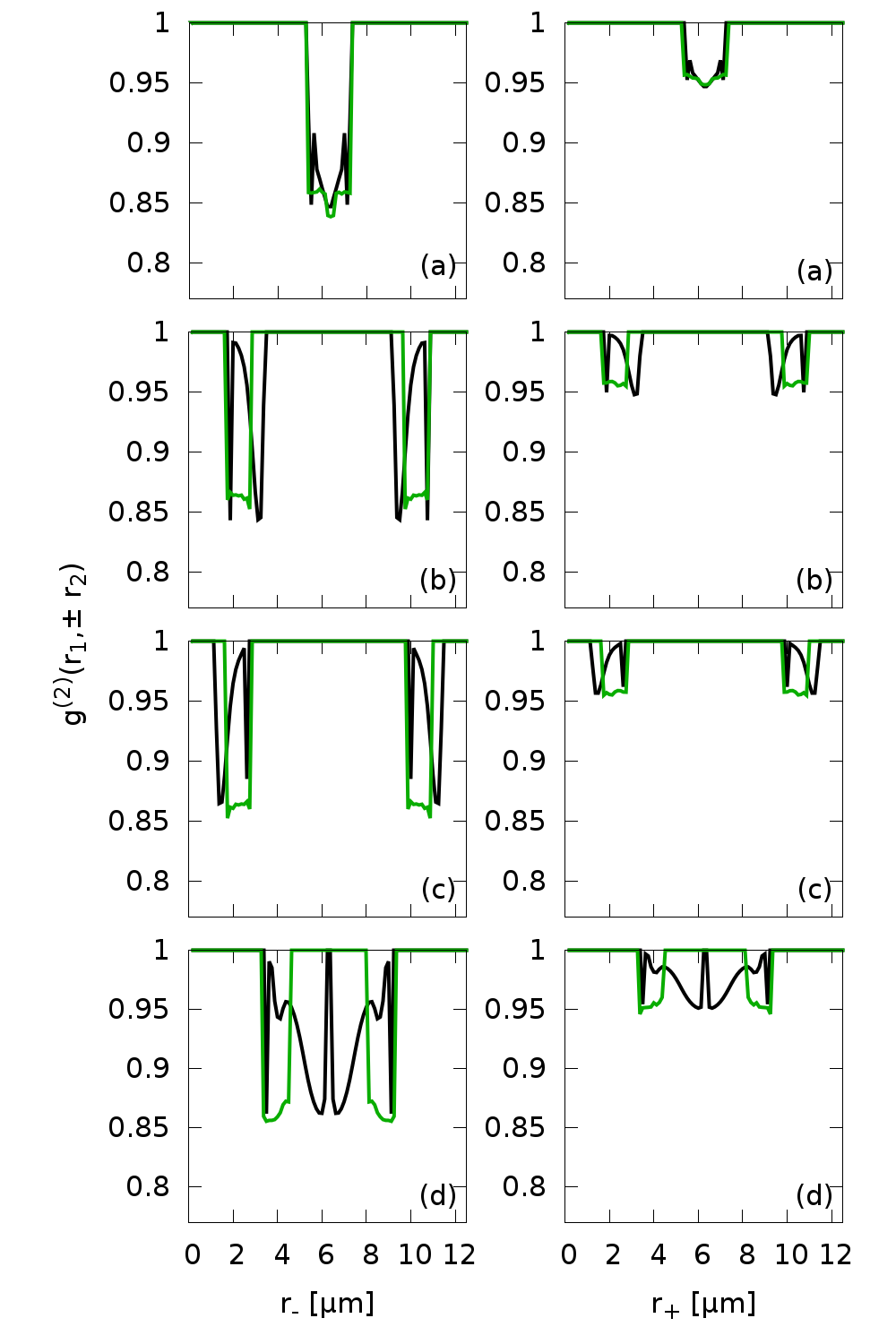}}
\caption{Associated one-dimensional diagonal cuts $g^2(r_{\pm},t)$ of the second order correlation function, exact (black) and MTEF (green), plotted at four time snapshots: (a) $t = 100 ~a.u.$, (b) $t = 600 ~a.u.$, (c) $t = 1200 ~a.u.$, (d) $t = 2100 ~a.u.$.}
\label{AB10}
\end{figure}

In Fig.~\ref{AB9} we show $g^2(r_1,r_2,t)$ for the two-photon emission process. The energy level spacing in the three-level truncation of the 1D soft-Coulomb hydrogen atom is uneven, such that the two emitted photons are of different frequencies. Hence, in contrast to the one-photon process, we expect to observe asymmetric features in the second order correlation function. In the exact result we observe that the vacuum state is locally disturbed by a structured, anti-bunched photon wavepacket. The fine, multi-lobed, spatial structure of the photon wavepacket is blurred into a single, rather narrow, feature in the MTEF result. However, MTEF dynamics indeed show the correct spatial asymmetry that is expected in $g^2(r_1,r_2,t)$. In the corresponding one-dimensional cuts of $g^{2}(r_{1},r_{2},t)$, shown in Fig.~\ref{AB10}, we show in further detail the comparison of MTEF dynamics and the exact results in this more complex two-photon case.

\section{Summary and Outlook}\label{sec:conc}

In this work we have adapted the multi-trajectory Ehrenfest method (MTEF) to simulate correlated quantum mechanical light-matter systems. We applied this mixed quantum-classical dynamics method, which is traditonally applied to electron-nuclear dynamics problems, to two and three level model QED cavity bound atomic systems, and in order to simulate observables and correlation functions for both the atomic system, and the photon field. We find that MTEF dynamics is able to qualitatively characterize the correct dynamics for one and two photon spontaneous emission processes in a QED cavity. However, MTEF dynamics does suffer from some quantitative drawbacks. Furthermore, we also observed that MTEF dynamics simulations can, in fact, capture some quantum mechanical features such as bound polariton states and second order photon correlations. Moreover, as experimental advances drive the need for realistic descriptions of light-matter coupled systems, trajectory-based quantum-classical algorithms emerge as promising route towards treating more complex and realistic systems. In particular, combining the \textit{ab initio} light-matter coupling methodology recently presented in Jest\"adt et. al. \cite{JRORA18} with our multi-trajectory approach, provides a computationally feasible way to simulate photon-field fluctuations and correlations in realistic three-dimensional systems. Work along these lines is already in progress. Furthermore, an alternative to the independent trajectory-based approach employed here is the conditional wavefunction approach, which allows one address nonadiabatic dynamics problems in complex systems with higher accuracy than MTEF dynamics\cite{Guille18}, and opens up an interesting potential route for mixed quantum-classical methods in correlated light-matter systems.

\section*{Acknowledgements}

We would like to thank Niko S\"akkinen and Johannes Flick for insightful discussions and acknowledge financial support from the European Research Council (ERC-2015-AdG-694097). AK acknowledges support from the National Sciences and Engineering Research Council (NSERC) of Canada.

\bibliographystyle{unsrt}
\bibliography{HKSRA18}

\end{document}